\documentclass{aastex62}

\received{March 8, 2019}
\revised{June 24,2019}
\submitjournal{ApJ}

\shorttitle{Gas scale heights and Kennicutt-Schmidt law}
\shortauthors{Wilson et al.}

\begin{document}

\title{The Kennicutt-Schmidt Law and Gas Scale Height in 
  Luminous and Ultra-Luminous Infrared Galaxies}

\author[0000-0001-5817-0991]{Christine D. Wilson}
\affil{Department of Physics and Astronomy, McMaster University, 1280 Main St. West,
Hamilton ON L8S 4M1 Canada}\email{wilsoncd@mcmaster.ca}

\author[0000-0002-1723-6330]{Bruce G. Elmegreen}
\affiliation{IBM T. J. Watson Research Center, 1101 Kitchawan Road, Yorktown
Heights, New York 10598 USA} \email{bge@us.ibm.com}

\author{Ashley Bemis} \email{bemisa@mcmaster.ca}
\affil{Department of Physics and Astronomy, McMaster University,
1280 Main St. West, Hamilton ON L8S 4M1 Canada}

\author{Nathan Brunetti} \email{brunettn@mcmaster.ca}
\affil{Department of Physics and Astronomy, McMaster University,
1280 Main St. West, Hamilton ON L8S 4M1 Canada}

\begin{abstract}
A new analysis of high-resolution data from the Atacama Large
Millimeter/submillimeter Array (ALMA) for 5 luminous or ultra-luminous infrared
galaxies gives a slope for the Kennicutt-Schmidt (KS) relation equal to
$1.74^{+0.09}_{\rm -0.07}$ for gas surface densities $\Sigma_{\rm
mol}>10^3\;M_\odot$~pc$^{-2}$ and an assumed constant CO-to-H$_2$ conversion factor. The velocity dispersion of the CO line, $\sigma_v$, scales approximately as
the inverse square root of $\Sigma_{\rm mol}$, making the empirical gas scale
height determined from $H\sim0.5\sigma^2/(\pi G\Sigma_{\rm mol})$ nearly constant,
150-190~pc, over 1.5 orders of magnitude in $\Sigma_{\rm mol}$. This constancy of
$H$ implies that the average midplane density, which is presumably dominated by
CO-emitting gas for these extreme star-forming galaxies, scales linearly with the
gas surface density, which, in turn, implies that the gas dynamical
rate (the inverse of the free-fall time) varies
 with $\Sigma_{\rm mol}^{1/2}$, thereby explaining most of the super-linear slope
in the KS relation. Consistent with these relations, we also find that the
mean efficiency of star formation per free-fall time is roughly constant, 5\%-7\%, and
the gas depletion time decreases at high $\Sigma_{\rm mol}$, reaching only $\sim
16$ Myr at $\Sigma_{\rm mol}\sim10^4\;M_\odot$ pc$^{-2}$. The
variation of $\sigma_v$ with $\Sigma_{\rm mol}$ and the constancy of $H$ are in
tension with some feedback-driven models, which predict $\sigma_v$
to be more constant and $H$ to be more variable. However, these results are consistent with
simulations in which large-scale gravity drives turbulence through a feedback
process that maintains an approximately constant Toomre $Q$
instability parameter. 
\end{abstract}


\section{Introduction} \label{sec:intro}

The Kennicutt-Schmidt (KS) relation describes the observed correlation between the star formation
rate per unit area, $\Sigma_{\rm SFR}$, and the surface density of gas, $\Sigma_{\rm
gas}$ and is a power law for the main disk regions of
spiral galaxies. Because star formation is expected to follow the gas, a slope close to unity,
as found for CO emission by \cite{bigiel08} and \cite{leroy08} or HCN emission by
\cite{gao04}, might not be surprising. However, star formation is a dynamical
process involving the rate of conversion of gas into stars, so a mass dependence
alone (as in the linear law) cannot be the full story. There has to be a time
component, and for gravitating systems, that means a volume density is
involved. 
The linear laws
only depend on the gas surface density, rather than the volume density,
so these laws presumably arise from
selection effects in surveys that observe sub-regions of gas at a characteristic
density, depending on the molecular transition used \citep{krumholzthompson07,nara08}.
The timescale is then the collapse time at that selected density, i.e., a constant.
In contrast, the total gas should have a continuum of densities that widely
participates in a gravity-driven condensation into dense clouds
\citep{elmegreen15,elmegreen18}. If the average density increases with $\Sigma_{\rm
SFR}$, then the KS slope will be steeper than linear, such as $1.4$ in the
observations by \cite{kennicutt98}, \cite{delosreyes2019}, and others.

For a disk with gas surface density $\Sigma_{\rm gas}$ and scale height $H$, the
average midplane gas density is $\rho_{\rm mid}=\Sigma_{\rm gas}/(2H)$, so the
observed total-gas slope of $\sim1.4$ can result from a gravity-driven model with a
rate $(G\rho_{\rm mid})^{0.5}$, provided that the disk scale height is about
constant \citep{madore77,larson88,elmegreen18}.  In the Milky Way, the thickness of
the molecular layer is indeed about constant inside the solar radius
\citep{heyer15}, but there is no direct view yet of the disk thickness in other
galaxies where the KS relation has been measured.

The KS relation for starbursts and (ultra)-luminous infrared galaxies (U/LIRGS) has
about the same $\sim1.5$ slope for CO as the total gas relation in galaxy disks
\citep{kennicutt98,gao04,krumholz12,gowardhan17,shi18}. This is presumably because
most of the gas in starbursts is dense enough to emit CO and that molecule is no
longer a sparse tracer subject to selection effects. The similar slope implies that
even with extremely high star formation rate densities, the balance between feedback
and self-gravity produces a vertical equilibrium with a relatively constant gas
thickness, i.e., much more constant than the range of surface densities.

The purpose of this paper is to examine more closely the KS relation in the
starburst regime and to estimate the disk thickness from the observed molecular gas
velocity dispersion and surface density. From these we determine the average
midplane density, free-fall time, gas consumption time, and efficiency per free-fall
time. The results confirm the super-linear KS slope found previously for starbursts,
and they also reveal a nearly constant disk thickness, confirming the most basic
model in which three-dimensional density primarily determines the rate at which gas
turns into stars
(\citealt{madore77,silk87,katz92,elmegreen94,elmegreen02,krumholz05,bacchini2019};
see review in \citealt{krumholz14}).

In what follows, Section \ref{sec:data} describes the observations and data
processing. Section \ref{sec:ks} derives the KS law, Section \ref{sec:h} determines
the disk scale heights, and Section \ref{section-times} derives the gas depletion time,
free-fall time, and efficiency per free-fall time. Section \ref{sec:disc} considers
our observations in the context of various theoretical predictions and Section
\ref{sec:conc} presents the conclusions.

\section{Observations and data processing} \label{sec:data}

To study the KS relation at high star formation rates, we searched the ALMA
archive for U/LIRGs for
which suitable
observations of the CO J=1-0 line were available
(Table~\ref{tab:obstable}). For each project, the
raw uv-data were calibrated using the scripts retrieved from the archive and the
CASA version used in the original calibration. 
All further processing was carried out in
CASA versions 5.0 to 5.4. Continuum subtraction was
performed on the uv-data using line-free channels. 
Cleaned image cubes were made using Briggs weighting
with robust=0.5 and channel widths of 20~km~s$^{-1}$ (26.4~km~s$^{-1}$
for NGC 3256, 40~km~s$^{-1}$ for
Arp 220). Continuum
images were made with the same weighting using the line-free channels. For three
galaxies where the CO and the continuum images used different ALMA data sets, a
common minimum uv-distance cutoff was used for both datasets and a taper was
applied to roughly match the resulting beams. Finally, the continuum
image and line cube were smoothed to have identical resolution. More details of
the image processing are given in \citet{wilson2019}.

\begin{table}
\renewcommand{\thetable}{\arabic{table}}
\centering
\caption{(Ultra-)Luminous Infrared Galaxies observed with ALMA}   \label{tab:obstable}
\begin{tabular}{lcccccc}
\tablewidth{0pt}
\hline
\hline
Galaxy & Distance\tablenotemark{a} & Map area\tablenotemark{b} & beam FWHM &
$\sigma_{\rm cont}$ & $\sigma_{\rm CO(1-0)}$ &
binned pixel \\ 
& (Mpc) & (kpc$^2$) & ($^{\prime\prime}$) & (mJy~beam$^{-1}$) & (mJy~beam$^{-1}$ km~s$^{-1}$) &  size (pc) \\
\hline
IRAS~17208-0014 & 182     & 1.3 & 0.5 & 0.05 & 0.18 & 397 \\ 
Arp~220 & 79     & 1.1 & 0.95$\times$0.60 & 0.10 & 0.18 & 345 \\ 
IRAS~13120-5453 & 134   & 3.4 & 1.1 & 0.08 & 0.14 & 650 \\
NGC~3256 & 44     & 5.2 & 2.2 & 0.05 & 0.12 & 512 \\ 
NGC~7469 & 66     & 1.6 & 0.95 & 0.015 & 0.043 & 418 \\ 
\hline
\end{tabular}
\tablenotetext{a}{From redshift (corrected to the 3K CMB reference frame) and assuming $H_o = 70.5$~km~s$^{-1}$~Mpc$^{-1}$. For NGC
  7469, SN Type Ia distance from Ganeshalingam et al. 2013.}
\tablenotetext{b}{The area of high signal-to-noise emission used in
  this analysis; the ALMA maps detect emission over  a larger area, especially in CO.}
\end{table}

We made integrated intensity (moment 0) and velocity
dispersion (moment 2) maps from the CO cubes using 3$\sigma$ and
4$\sigma$ cutoffs, respectively, and limiting the range to the
channels containing CO emission. The primary beam correction was applied to the
CO integrated intensity and continuum images.
We make no correction for channelization effects in the moment 2 maps
\citep[cf.][]{sun2018}; this will cause $\sigma_v$ to be slightly overestimated,
but we estimate the effect is at most 12\% for the narrowest lines in our data.
We also use various combinations of these three maps to calculate
the dynamically-derived quantities described in \S\ref{sec:results}.
All images were then
rebinned so that individual pixels would be
approximately the size of the beam and therefore essentially independent.
We calculated uncertainty images for
the continuum and integrated intensity maps that included both the 5\%
absolute calibration uncertainty and the statistical measurement uncertainty.
We propagate the uncertainties through the formula used to calculate
the moment 2 map to obtain an equation for the uncertainty in the moment 2 map,
$(\sigma_I/I)(N_{chan}\Delta v_{chan})^2/17.68/\sigma_v$, where
$I$ and $\sigma_I$ are the CO J=1-0 integrated intensity and its
uncertainty, 
$N_{chan}\Delta v_{chan}$ is the velocity range used to
calculate the moment maps,
and $\sigma_v$ is the velocity dispersion from the moment 2 map.
Only binned pixels with a signal-to-noise greater than 4 in all three
images are included in our analysis below.

Our relatively small pixel sizes mean that 
the contribution to $\sigma_v$ from systematic velocity gradients, e.g., shear,
inside each resolution element is small. 
We estimated this beam-smearing effect for IRAS 13120, which is the
galaxy with the lowest spatial resolution in pc and so is likely to be
the most affected. We removed the uvtaper to produce moment 1
(velocity field) and moment 2 
(velocity dispersion) maps at a resolution of $0.56^{\prime\prime}$ to
compare with our fiducial maps. We combined the moment 1 and moment 2
maps at both resolutions to determine 
corrected velocity dispersion maps from the quadratic differences
between the mean 
value for the velocity dispersion and
the standard deviation of the velocity field across each beam.
At both resolutions, this procedure produced only  
a small decrease in $\sigma_v$  (2-7\%) compared to $\sigma_v$ measured directly from
the moment 2 map. 
We also compared
the velocity dispersion averaged over $1.1^{\prime\prime}$ pixels on maps at the two
different resolutions. On average, the velocity dispersion in the
$1.1^{\prime\prime}$ maps is 18\% larger than the value from the
$0.56^{\prime\prime}$ maps. Putting these two results together, we estimate that the
velocity dispersion in a typical pixel is overestimated by at most
20\%. We note that the very central
pixel towards each galaxy nucleus, which is also typically the pixel
with the highest gas surface density, might be more strongly affected
by beam smearing. A pervasive overestimate of $\sigma_v$ by 20\% from velocity
gradients inside the beam will not affect the slopes of the various 
scaling relations
discussed in the next section.

\section{The Kennicutt-Schmidt law and dynamically-derived quantities
} \label{sec:results}

\subsection{The K-S relation at high surface densities}
\label{sec:ks}

\begin{figure}
\plottwo{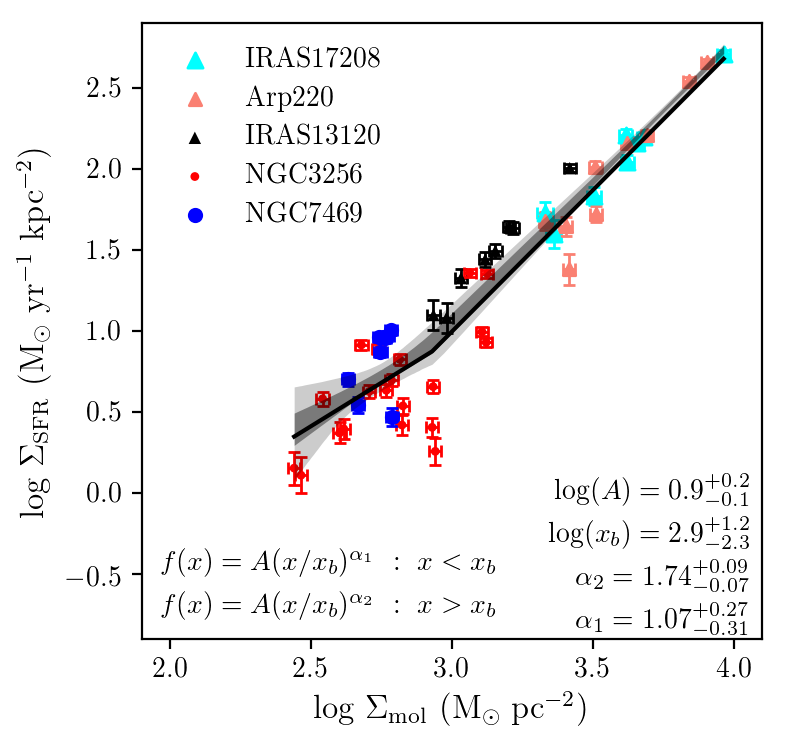}{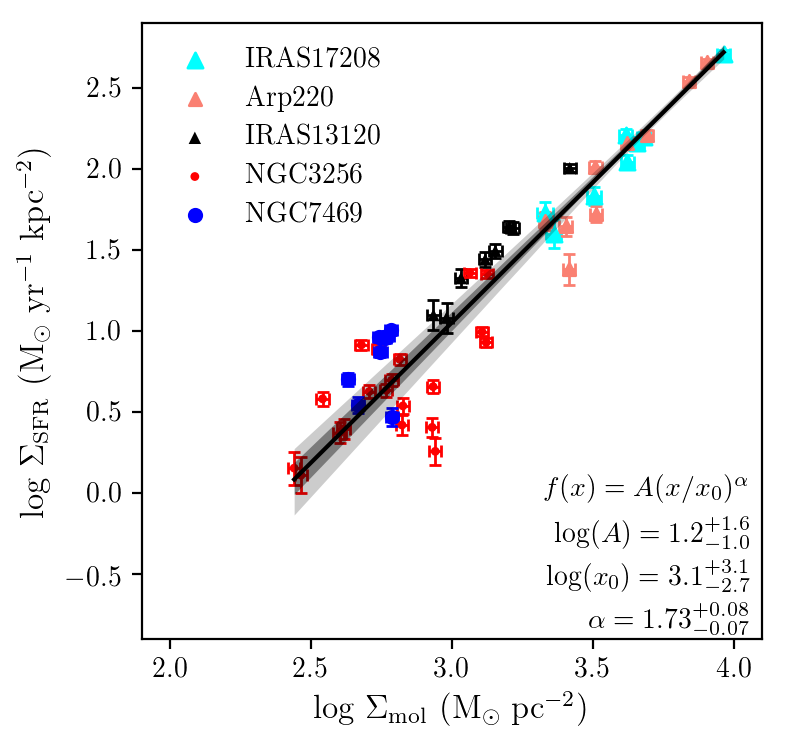}
\caption{The resolved Kennicutt-Schmidt relation for five U/LIRGs
  fit with (left) a double power law and (right) a
  single power law. The conversion factor, $X_{\rm CO}$, is assumed to be
  constant. Note that the slope of the KS relation is substantially
  greater than 1 at high gas surface densities. \label{fig:ksplot}}
\end{figure}

We adopt the U/LIRG value for the CO-to-H$_2$ conversion factor \citep{downes1998}
and include a factor of 1.36 for helium
to convert the CO
integrated intensities, $I_{\rm CO(1-0)}$ in K~km~s$^{-1}$, to observed molecular
gas surface densities, $\Sigma_{\rm mol}$ in M$_\odot$~pc$^{-2}$, $\Sigma_{\rm
mol} = 1.36 \Sigma_{\rm H_2} = 1.088 I_{\rm CO(1-0)}$.
We make no correction for
the (poorly constrained)
inclination of these disturbed systems, which will cause
$\Sigma_{\rm mol}$ (and $\Sigma_{\rm SFR}$) to somewhat overestimate the true surface density
perpendicular to the disk.
These surface densities exceed 100~M$_\odot$~pc$^{-2}$
and so we ignore any contribution from atomic
gas.
The expected high dust extinction means we need a
star formation rate tracer that minimizes the effect of dust while still
providing arcsecond-scale resolution. The one star formation rate
tracer that meets both these requirements is the radio continuum
\citep{murphy2011}.
We exclude the
nucleus of NGC~7469 from our analysis as this galaxy
contains a strong active galactic nucleus (AGN) that  contributes a
significant fraction of the radio continuum emission.

In such gas-rich systems, thermal emission from dust can also contribute
to the 93-106~GHz emission. We estimated the dust contribution
by comparing published fluxes or ALMA images at
330-350~GHz (230~GHz for NGC~3256) with our continuum images. We
find that dust contributes on average 10\% of the emission at 93
GHz (15\% at 106~GHz); the relative contribution at these two
frequencies is consistent with  a dust emissivity index $\beta
\sim 1.5-1.8$. The one exception is the western nucleus of
Arp~220, where the dust
contributes 40\% of the flux \citep{sakamoto2017}.
We correct our measured continuum fluxes by these various factors to
remove the contribution from dust emisson before calculating the star
formation rate.

The dust-corrected 93-106~GHz emission from these galaxies likely contains a mixture
of thermal (free-free) and non-thermal (synchrotron) emission. Assuming an
excitation temperature of $10^4$~K and a non-thermal spectral index $\alpha_{\rm
NT}=0.83$ \citep{murphy2011}, non-thermal
  emission should contribute $\sim$25\% of the
total emission at these frequencies, e.g. a thermal:non-thermal ratio
of 3:1. 
We therefore calculate the star formation
rate surface density, $\Sigma_{\rm SFR}$, using the
thermal-only formula from \citet{murphy2011}.
[For emission at 93~GHz,
the thermal-only equation gives a $\Sigma_{\rm SFR}$ that is 24\% larger than the
value obtained with the standard thermal+non-thermal equation from
\citet{murphy2011}.] 
However, it is possible for the thermal radio emission to be reduced
if some of the ionizing photons are directly absorbed by dust \citep{murphy2011}.
Such absorption is difficult to quantify but could be an important process in
these extreme systems. For example, \citet{sakamoto2017} estimate the
majority of the 106~GHz 
emission in Arp~220 is non-thermal. 
For NGC~7469, we have compared an archival 8~GHz image with our 93 GHz
image, which suggests
that $\sim$50\% of the 93~GHz emission is non-thermal, 
 e.g. a thermal:non-thermal ratio of 1:1 across the inner disk.
Adjusting the standard thermal+non-thermal equation from
\citet{murphy2011} by assuming that dust absorption suppresses the 
thermal emission by a factor of 3 would double
$\Sigma_{\rm SFR}$ compared to the values used here.

Figure~\ref{fig:ksplot} shows the resolved KS relation
for
our sample. We  fit the relation with a double power
law using the Astropy Modeling package with the break point location
as a free parameter, and
bootstrapped 10000 times to get the fit parameters and uncertainties. We
find a slope of $1.74^{+0.09}_{\rm -0.07}$ in the high surface
density regime, with some indication of a shallower slope
and increased scatter at surface densities below $\sim 1000$~M$_\odot$~pc$^{-2}$.
A single power law fit to all of the data yields a nearly identical slope ($1.73^{+0.08}_{\rm -0.07}$).

This steep power-law slope differs from the usual KS relation derived for CO
emission, which is linear for local galaxies where $\Sigma_{\rm CO}$ tends to be
less than several hundred M$_\odot$~pc$^{-2}$ \citep{bigiel08,leroy13}. For U/LIRGs
in general, CO is a good measure of total gas surface density because most of the
gas exceeds the threshold for CO emission. The transition from a linear CO law to a
steep CO law at high $\Sigma_{\rm gas}$ was predicted in \citet{elmegreen15}.
\citet{gao04} and \citet{shi18} also find a relatively steep CO law at high surface
densities.

\cite{narayanan12} have suggested that the CO-to-H$_2$ conversion factor, $X_{\rm CO}$,
decreases with increasing CO intensity as 
\begin{equation}
X_{\rm CO}={\rm min}(4,6.75/W_{\rm CO}^{0.32})\times10^{20}\;\;{\rm cm}^{-2}({\rm K\; km\;s}^{-1})^{-1}
\label{eq:xco}
\end{equation}
where $W_{\rm CO} = I_{\rm CO(1-0)}$; this equation does not include the
  factor for helium.
\citep[A similar result based more on star formation history than the instantaneous
rate was suggested by][from the simulation of a merger.]{renaud19} This
 inverse dependence on $W_{\rm
  CO}$ would steepen the KS relation. If we
write $X_{\rm CO}\propto W_{\rm CO}^{-x}$, then a KS relation like $\Sigma_{\rm
SFR}\propto \Sigma_{\rm gas}^y$ with an assumed constant $X_{\rm CO}$ converts to a
steeper KS relation, $\Sigma_{\rm SFR}\propto\Sigma_{\rm gas}^{y/(1-x)}$, with a
variable $X_{\rm CO}$. Taking $y=1.74$ and $x=0.32$, the revised slope
would be $2.6$. In
what follows, we proceed with the assumption of a constant $X_{\rm CO}$ to
facilitate comparisons with other studies, but we note the effect that equation
(\ref{eq:xco}) would have on the slopes of the other relationships derived below.

\subsection{Gas scale height}
\label{sec:h}

The equilibrium thickness of the gas disk in a galaxy depends on the balance between
confining pressure from gravity and uplifting pressures from gas motions, magnetic
fields, and cosmic rays \citep[e.g.,][]{deavillez05,girichidis16,hill18}. The
gravitational forces are proportional to the total mass surface density inside the
gas layer, $\Sigma_{\rm total,GL}$, plus vertical components of galactic gravity
from remote regions. The relevant total surface density consists of
gas plus the stars
and the dark matter that reside inside the gas layer. Upward magnetic pressure including
field line tension depends on the difference between $(B^2-2B_z^2)/8\pi$ at the
midplane and at the gas scale height, where $B$ is the total field strength and
$B_{\rm z}$ is the vertical component \citep{boulares90};  cosmic ray pressure
depends on the analogous difference for cosmic rays. \cite{ostriker10} suggest that
the magnetic and cosmic ray contributions to the pressure gradient are much smaller
than the turbulent pressure gradient because magnetic fields and cosmic rays extend
to much greater heights than gas. \cite{kim15} model a shearing box of magneto-hydrodynamic
turbulence with star formation feedback and determine that the vertical magnetic
pressure gradient contributes an additional disk support that is $\sim0.3$ times the
support from turbulent and thermal pressures. Denoting the average ratios of these
pressure differences to the gas pressure by the constants $\alpha\sim0.3$ for the
magnetic to turbulent plus thermal support ratio and $\beta\sim0$ for
the cosmic ray
to turbulent plus thermal support ratio \citep{parker66}, the upward pressure is
$P_{\rm ISM}=\rho_{\rm mid}\sigma_{\rm v}^2(1+\alpha+\beta)$ for average midplane
gas density $\rho_{\rm mid}$ and combined thermal and turbulent velocity dispersion
in the vertical direction, $\sigma_{\rm v}$.

The confining pressure from the disk is $P_{\rm grav}=0.5\Sigma_{\rm gas}{\bar g}$,
where $0.5\Sigma_{\rm gas}$ is the gas surface density in one-half the layer and
${\bar g}$ is the average gravitational acceleration to the midplane in that half.
The gravitational acceleration comes from Poisson's equation, $\nabla \bullet g=4\pi
G\rho_{\rm total,GL}$, with Gauss's solution, giving $g=2\pi G\Sigma_{\rm total,GL}$
at the effective top of the gas layer, i.e., at one scale height, $H$. For a
constant vertical velocity dispersion, $g$ increases approximately linearly with
height, so ${\bar g}=0.5g$ and $P_{\rm grav}=0.5\pi G\Sigma_{\rm gas}\Sigma_{\rm
total,GL}$. These equations also come from exact solutions to the vertical
equilibrium of an isothermal layer \citep{spitzer42}. Setting $P_{\rm ISM}=P_{\rm
grav}$ in equilibrium and rearranging gives the gas disk half-thickness,
\begin{equation}
H={{\Sigma_{\rm gas}}\over{2\rho_{\rm mid}}}={{\sigma_{\rm v}^2(1+\alpha+\beta)}\over
{\pi G \Sigma_{\rm total,GL}}}.
\label{eq:basicH}
\end{equation}

For some applications, it is important to consider vertical forces from
additional mass outside the disk region. One example of such a force is the
vertical component of the three-dimensional gravitational acceleration toward the
inner part of the galaxy, which contains the total mass that also gives the
rotation curve at velocity $v_{\rm rot}(R)$ for galactocentric radius $R$. At
height $H$, this is the geometric fraction $H/R$ of the total acceleration,
$GM_{\rm galaxy}/R^2$ for galaxy mass $M_{\rm galaxy}$ inside $R$. The average
perpendicular acceleration in the gas layer is about half of this, ${\bar g}_{\rm
galaxy}=0.5GM_{\rm galaxy}H/R^3$. The ratio of this to the pure disk component is
\begin{equation}
{{{\bar g}_{\rm galaxy}}\over{{\bar g}}}=0.5\left({{M_{\rm galaxy}}\over{M_{\rm disk,GL}}} \right)
\left(H\over R\right),
\label{gequation}
\end{equation}
where $M_{\rm disk,GL}=\pi R^2\Sigma_{\rm total,GL}$ is the effective disk mass
inside the vertical thickness of the gas layer out to radius $R$, ignoring
gradients in surface density. This ratio enters Equation (\ref{eq:basicH}) as
$(1+{\bar g}_{\rm galaxy}/ {\bar g})$ multiplying the surface density,
$\Sigma_{\rm total,GL}$, in the denominator.

\begin{figure*}
\gridline{\fig{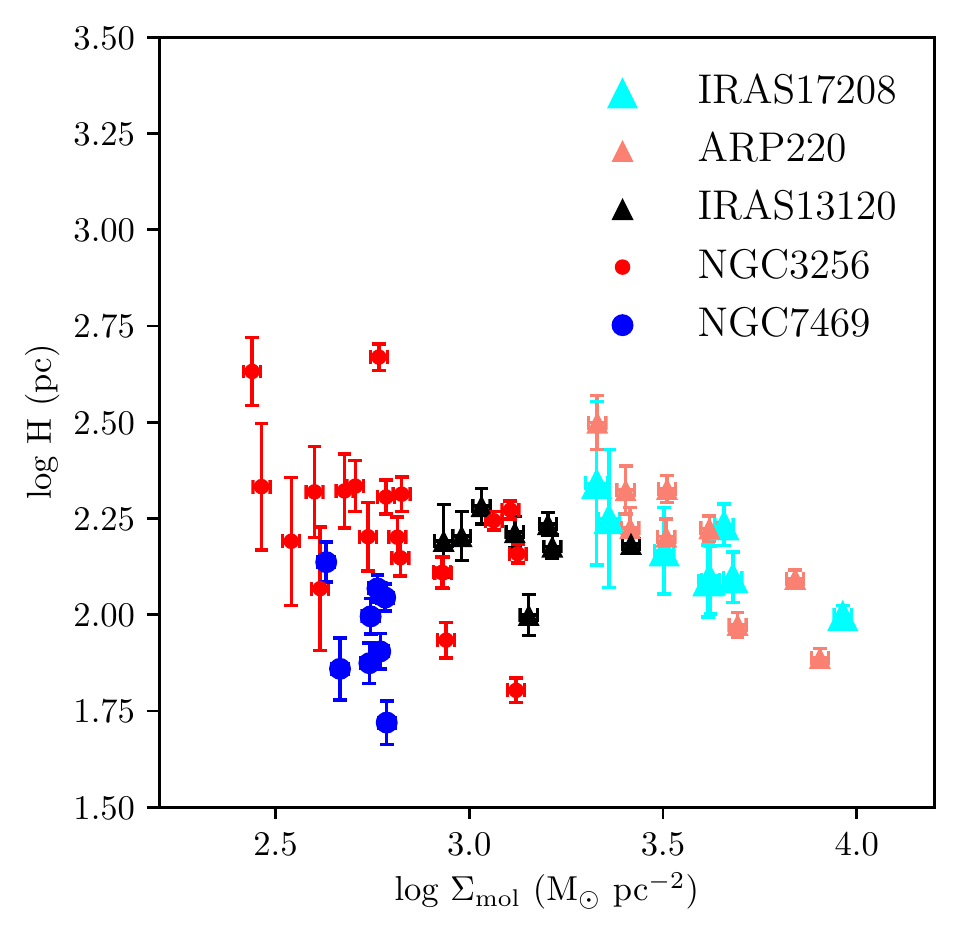}{0.4\textwidth}{(a)}
          \fig{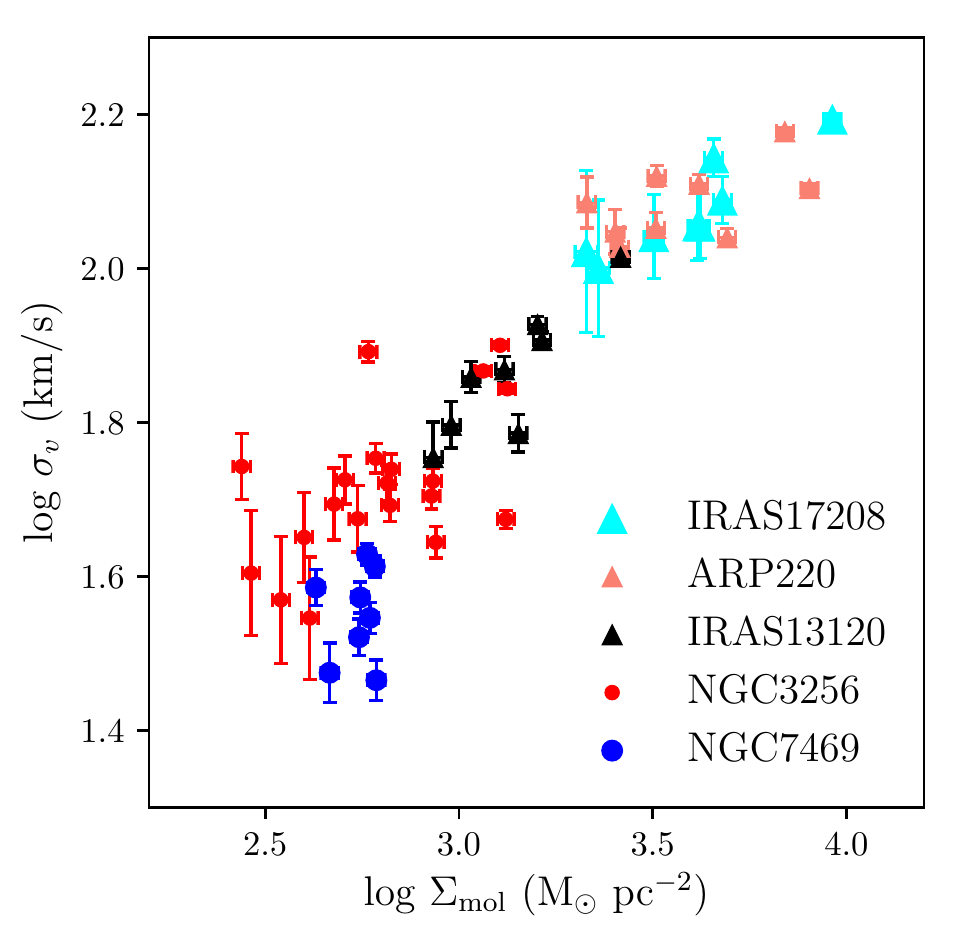}{0.4\textwidth}{(b)}
          }
\caption{(a) The scale height of the molecular gas (Equation~\ref{eqH})
plotted as a function of the molecular gas surface density.
$H$ is surprisingly constant over a wide range in surface density.
(b) The velocity dispersion of the molecular gas is plotted as a function of the
molecular gas surface density; the slope is approximately 0.5.
\label{fig:2panel}}
\end{figure*}

Writing $\mu=M_{\rm disk,GL}/M_{\rm galaxy}<1$ as the ratio of masses and
$\xi=\sigma_v^2(1+\alpha+\beta)/v_{\rm rot}^2<1$ as the squared ratio of the gas
velocity dispersion supplemented by magnetic and cosmic ray pressures to the
galaxy rotation speed, Equation (\ref{eq:basicH}) including this additional force
of gravity becomes
\begin{equation}
\left({H \over R}\right)\left(1+{{H}\over{2\mu R}}\right)={\xi \over \mu}
\end{equation}
which has the solution
\begin{equation}
{H\over R}=\left(\mu^2+2\xi\right)^{1/2}-\mu.
\end{equation}
Sample values for a galaxy disk might be $\mu\sim0.5$ and $\xi\sim0.1$; for such
$\xi<<\mu$, $H/R\sim\xi/\mu\sim0.2$ in this case. Then ${\bar g}_{\rm galaxy}/{\bar
g}\sim\xi/(2\mu^2)\sim0.2$  is a small correction to Equation (\ref{eq:basicH}). For
ULIRG centers, the disk may dominate giving $\mu\sim1$; also, for
$\sigma_v/v_{rot}\sim 0.3$ (as in IRAS13120-5450), $\xi\sim0.12$ (assuming
$\alpha+\beta=0.3$). Then $H/R\sim0.11$ and ${\bar g}_{\rm galaxy}/{\bar
g}\sim0.05$.

Combining these terms, equation (\ref{eq:basicH}) becomes
\begin{equation}
H={{\sigma_{\rm v}^2}\over
{\pi G \Sigma_{\rm gas}}}\times\left({1+\alpha+\beta}\over{1+{\bar g}_{\rm galaxy}/{\bar g}}\right)
\times\left({\Sigma_{\rm gas}}\over{\Sigma_{\rm total,GL}}\right).
\label{eq:newH}
\end{equation}
In the solar neighborhood, the stellar midplane density is
$0.043\pm0.004\;M_\odot$~pc$^{-3}$, the dark matter density is
$0.013\pm0.003\;M_\odot$~pc$^{-3}$,  and the gas density is
$0.041\pm0.004\;M_\odot$~pc$^{-3}$; \citep{mckee15}. Thus locally, $\Sigma_{\rm
gas}/\Sigma_{\rm total,GL}\sim0.42$. In U/LIRGs, the stellar and gas densities might
be much higher than the dark matter density because of torques that drive the disk
mass inward, and then $\Sigma_{\rm gas}/\Sigma_{\rm total,GL}\sim0.5$. In gas-rich
galaxies at high redshift, the dark matter and gas surface densities could be
comparable and the stellar surface density slightly smaller, making the ratio around
0.5 again.  We discussed above the term from remote gravity, concluding that ${\bar
g}_{\rm galaxy}/{\bar g}\sim0.05$ to 0.2 for conditions representative of our
galaxies. For the magnetic and cosmic ray contributions to supporting pressure, we
follow the suggestion in \cite{kim15} that $\beta+\alpha\sim0.3$. Thus the second
and third terms in equation \ref{eq:newH} combine to give a factor of $\sim0.5$ and
we write for our highly molecular galaxies,
\begin{equation}
H \simeq 0.5 {{\sigma_v^2}\over {\pi G \Sigma_{\rm mol}}}.
\label{eqH}
\end{equation}

We consider the value of $H$ in equation \ref{eqH} to be an {\it empirical} scale
height because the relevant quantities are directly observable for a moderately
inclined galaxy. The approximations discussed above suggest there might be
$\sim50$\% variations from region to region in normal and starburst disks. More
detailed discussions of vertical equilibrium are in \cite{narayan02}, with further
applications in, for example, \cite{banerjee11}, \cite{elmegreen11},
\cite{elmegreen15a}, \citet{benincasa2016}, and \citet{bacchini2019}.


Equation~\ref{eqH} was used to calculate pixel-by-pixel maps of the distribution of
$H$ for each galaxy, where $\sigma_v$ is the velocity dispersion of the molecular gas as measured by the
moment 2 map.
Figure~\ref{fig:2panel}(a) shows that $H$ is relatively constant across our sample.
The mean values range from $150 \pm 15$~pc in NGC~7469 to $190 \pm 20$~pc in
NGC 3256. These scale heights are a factor of $\sim 2$ larger than the molecular gas
scale heights derived for 6 spiral galaxies where H$_2$ dominates in the central kpc
\citep{bacchini2019}; however, this factor of 2 is much less than the $10^4$ range
in gas surface density in the two samples combined.

Figure~\ref{fig:2panel}(b) plots the velocity dispersion versus the molecular
surface density. The dispersion, $\sigma_{\rm v}$, increases as
roughly the square root of
$\Sigma_{\rm mol}$ for the combined sample and also for different positions inside
each galaxy, except for NGC 7479 where the range in surface density is small. This
increase is consistent with the constancy of $H$, considering equation (\ref{eqH}).
The velocity dispersions range from 30 km s$^{-1}$ to 160 km s$^{-1}$, and are much
higher than in normal galaxy disks. Given the highly concentrated star formation and
high orbital speeds in U/LIRGs, it is not surprising that their gas velocity
dispersions would be much higher and their gas disks slightly thicker than in more
quiescent spiral galaxies.

The three galaxies in Figure~\ref{fig:2panel} with significant ranges
of gas surface density (IRAS~17208, Arp~220, and NGC 3256) show a
trend for $H$
to decrease slightly with increasing $\Sigma_{\rm mol}$. For the two
ultraluminous galaxies, $\sigma_{\rm v}$ may also be 
slightly more constant with $\Sigma_{\rm mol}$. These galaxies have
the highest $\Sigma_{\rm mol}$, greater than $2000\;M_\odot$ pc$^{-2}$ and also the
highest $\sigma_{\rm v}$, exceeding 100 km s$^{-1}$. These linewidths are getting
close to the rotation speeds of galaxies, and the excessively high
surface densities suggest
overlap or strong shock regions in these merging systems. Significant deviations
from the plane-parallel model of equilibrium vertical support should be expected.
Still, their average empirical thicknesses are comparable to those in the other
galaxies. 

If $X_{\rm CO}$ decreases with the integrated CO line as $W_{\rm CO}^{-x}$ for
$x=0.32$ \citep{narayanan12}, then $H$ versus $\Sigma_{\rm mol}$ and $\sigma_{\rm v}$
versus $\Sigma_{\rm mol}$ would both become steeper. 
The calculation of $H$ shown in Figure~\ref{fig:2panel} assumes $X_{\rm CO}$ is constant, so that a constant
$H$ means that $\sigma_v^2/W_{\rm CO}$ is constant. With a variable
$X_{\rm CO}$, $\Sigma_{\rm mol} \propto W_{\rm CO}^{1-x}$ and so
we
should have plotted $\sigma_{\rm v}^2/W_{\rm CO}^{1-x}$.
Given that
$\sigma_v^2/W_{\rm CO}$ is constant, this new $H$ would be
proportional to $W_{\rm CO}^x$, 
which means that $H\propto \Sigma_{\rm mol}^{x/(1-x)}\propto
\Sigma_{\rm mol}^{0.47}$. Similarly, $\sigma_{\rm v}$ would be proportional to 
$\Sigma_{\rm mol}$ to the power $0.5/(1-x)=0.74$ instead of 0.5.

\subsection{Gas depletion time, free-fall time, and efficiency per free-fall time}\label{section-times}

\begin{figure*}
\gridline{\fig{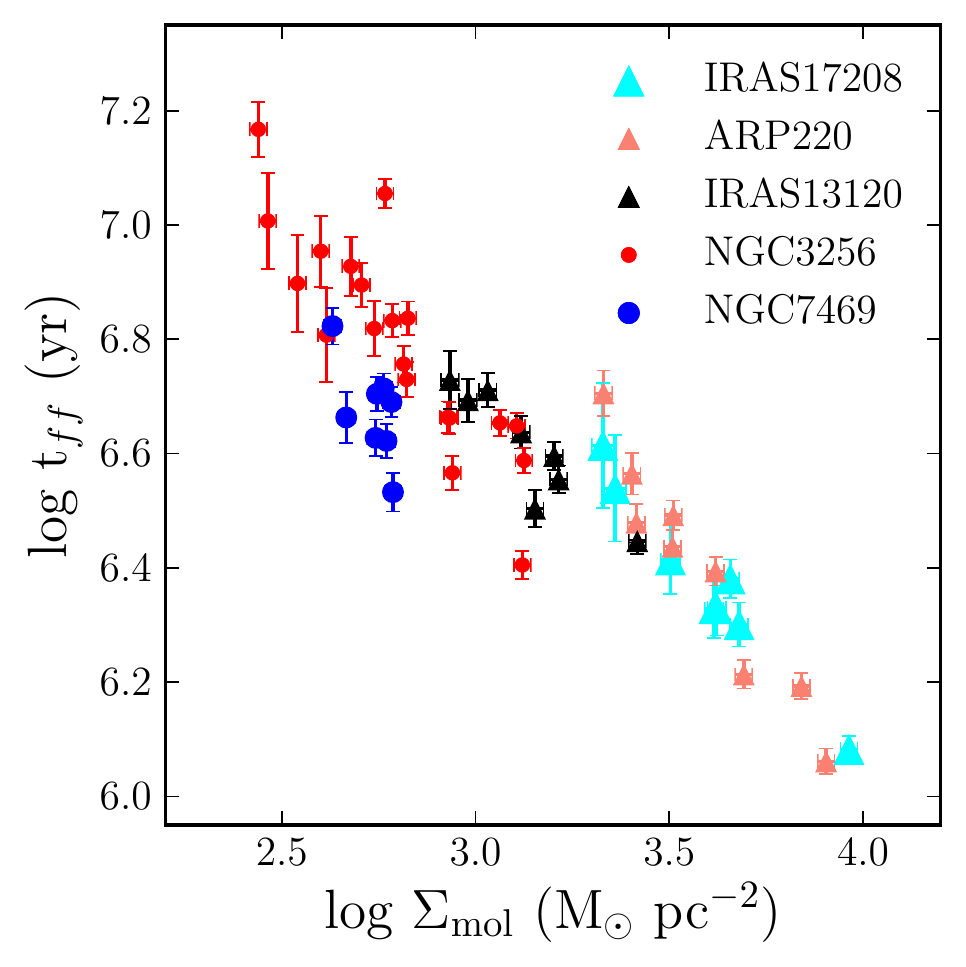}{0.3\textwidth}{(c)}
         \fig{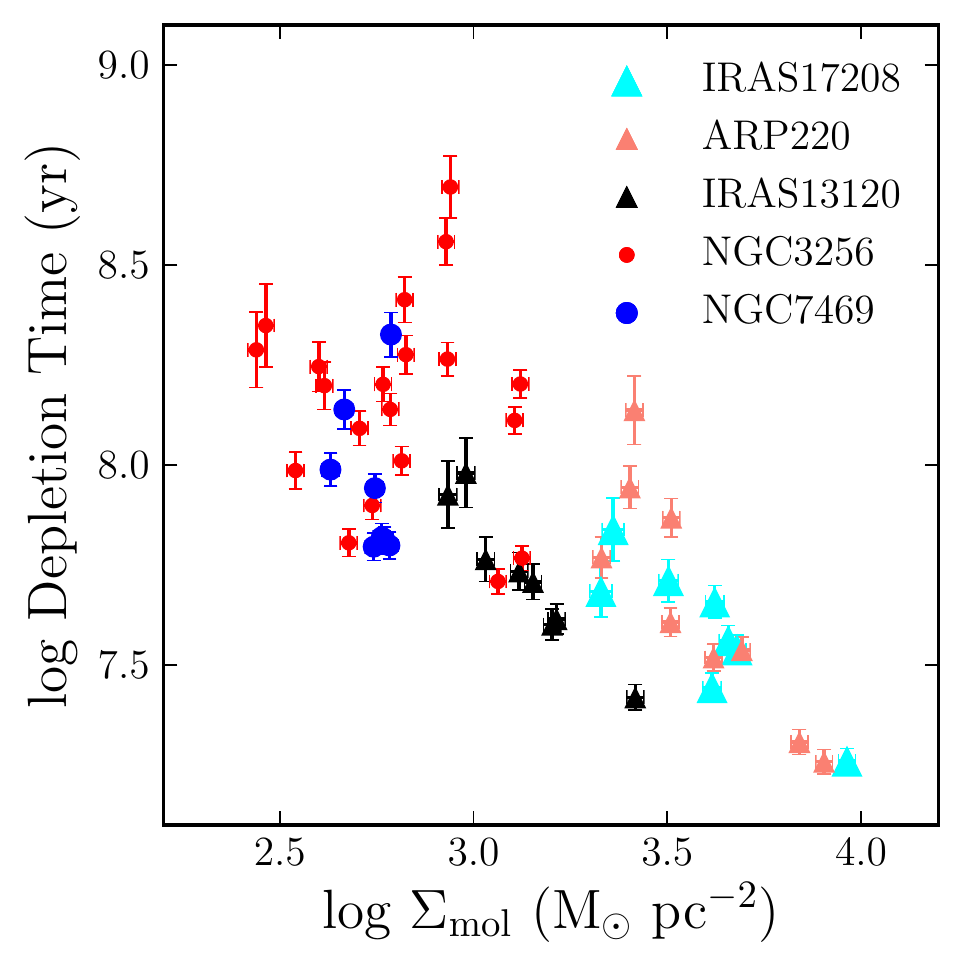}{0.3\textwidth}{(d)}
          \fig{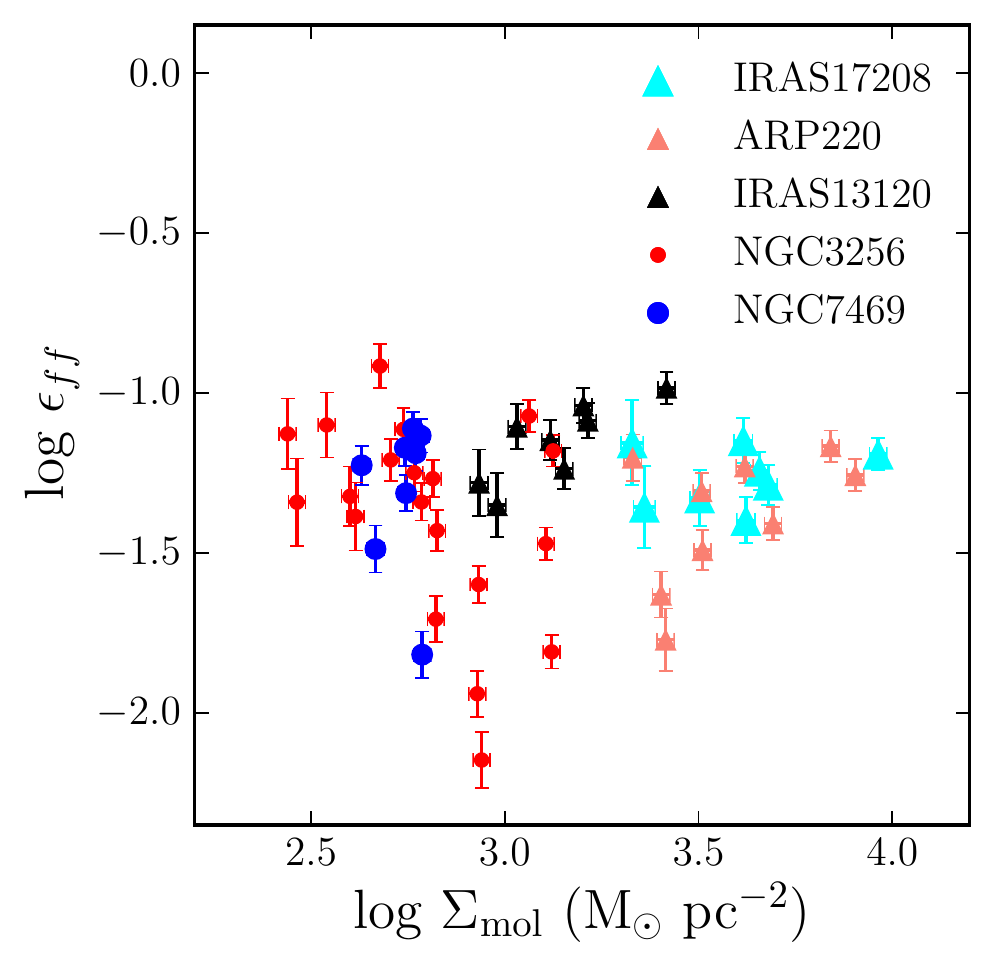}{0.3\textwidth}{(e)}
          }
\caption{(a) The free-fall time in the molecular gas plotted as a function of the
moleculr gas surface density. The dependence on $\sim1/\sqrt{\Sigma_{\rm mol}}$
is a direct consequence of the KS relation (Figure~\ref{fig:ksplot}) and
the constant gas scale height.
(b) The instantaneous gas depletion time shows a similar dependence on
the molecular gas surface density as the free-fall time.
(c) The star formation efficiency per free-fall time shows no
significant trend with gas
surface density and mean values are significantly higher than in spiral galaxies.
\label{fig:3panel}}
\end{figure*}

We now estimate the mid-plane gas density from
$\rho_{\rm mid} = \Sigma_{\rm mol}/(2H)$ and calculate the free-fall time,
$t_{\rm ff}$, via
\begin{equation}
t_{\rm ff} = {\sqrt{{3\pi}\over{32 G \rho_{\rm mid}}}} =
{\sqrt{3}\over{4 G}} {\sigma_v \over {\Sigma_{\rm mol}}}.
\label{eqtff}
\end{equation}
Figure~\ref{fig:3panel} shows that the free-fall time decreases as the gas surface
density increases, which is a consequence of the nearly constant scale height such
that $\rho_{\rm mid} \propto \Sigma_{\rm mol}$. It is also striking that the
free-fall times at the highest surface densities are extremely short ($< 2$ Myr). If
we adopt our beam size as an alternative measure of the extent of the gas emission
along the line of sight \citep{utomo2018}, then the mid-plane density would be
reduced by factors of only 1 to 2.3 and the free-fall times would be increased by
factors of only 1 to 1.5 for our sample. The variable X(CO) from
Eq. \ref{eq:xco} would have only a small effect on the observed trend.

Larger possible errors for the most extreme regions might arise from the ${\bar
g}_{\rm galaxy}/{\bar g}$ term as a scale height correction (Eq. \ref{gequation}) in
the bulge or nuclear regions where the stellar density might be high, and by
ignoring excessive magnetic and cosmic ray pressures. For example, the maximum
average midplane density implied from our analysis is $n_{\rm H_2} \sim 600$
cm$^{-3}$. This is much smaller than the mean densities implied by $\sim 120$~pc
resolution observations of Arp~220, which are $>(2-9)\times 10^4$ cm$^{-3}$ towards
the two nuclei \citep{wilson2014}. In the eastern nucleus,
where $\Sigma_{\rm mol}\sim 7
\times10^4\;M_\odot$ pc$^{-2}$ \citep{wilson2014},
\citet{rangwala15} have modeled the CO
emission as a turbulent rotating disk and estimate the velocity
dispersion to be 85 km s$^{-1}$. The high surface density combined
with a relatively modest velocity dispersion implies an empirical
scale height of just 4 pc, which seems to be unphysically
small. Clearly the simple formalism of Eq.~\ref{eqH} breaks down in
this regime; additional upward
force would have to come from additional magnetic and cosmic ray
pressures.
A more accurate calculation would correct for the inclination of the
disk, which
\citet{barcos-munoz15} estimate
to be 53.5 degrees. In addition, the surface density in \citet{wilson2014}
  is in fact an upper limit obtained by assuming
  the gas mass is equal to the dynamical mass. If we assume instead that
  the gas makes up 50\% of the total mass in this region and correct
  for inclination and helium, the face-on surface density becomes
  $\Sigma_{\rm mol} = 2.2 \times 10^4$ M$_\odot$ pc$^{-2}$. Correcting
  the equation for $H$ to use $\mu = 0.5$, $\sigma_v/v_{rot}\sim 1$
  \citep{rangwala15}, and $\alpha = \beta = 1$ to include increased
  magnetic and cosmic ray pressure, we obtain $H \sim 25$
  pc. This value for $H$ in turn implies a mid-plane density of
  $1.3\times 10^4$ cm$^{-3}$ and free-fall time $t_{\rm ff}$ of just
  0.27 Myr. We do not have an easy measure of $\Sigma_{\rm SFR}$ on
  this same scale; if we use our KS fit to estimate it from
  $\Sigma_{\rm mol}$, we obtain $\Sigma_{\rm SFR} = 2500$ 
M$_\odot$ kpc$^{-2}$ yr$^{-1}$. This in turn gives a gas depletion
time
$t_{\rm dep} = 8.8$ Myr and efficiency per free fall time
$\epsilon_{\rm ff} = 0.031$ (see below).


We calculate the instantaneous gas depletion time, $t_{\rm dep}$ from
\begin{equation}
t_{\rm dep} = {\Sigma_{\rm mol}\over \Sigma_{\rm SFR}}.
\label{eqtdep}
\end{equation}
Figure~\ref{fig:3panel} shows that, like the free-fall time, the gas depletion time
decreases as the gas surface density increases, although with increased scatter at
lower surface densities. This decrease is a natural result of the power-law slope of
1.7 seen in the KS relation in Figure~\ref{fig:ksplot}. It implies that starbursts
are truly bursty: they cannot sustain their high star formation rates for very long
unless gas accretes into the starburst region at an equally high rate.
The variable X(CO) from
Eq. \ref{eq:xco} would make the depletion time fall off more rapidly
with increasing surface density.

The decrease in $t_{\rm dep}$ toward high $\Sigma_{\rm mol}$ is not new. For
example, \cite{utomo17} recently found a similar result for normal galaxies that
$t_{\rm dep}$ decreases in galaxy centers where $\Sigma_{\rm mol}$ is higher.  Such
a trend is expected for a superlinear KS relation $\Sigma_{\rm
SFR}\propto\Sigma_{\rm gas}^y$ as the ratio $\Sigma_{\rm gas}/\Sigma_{\rm
SFR}\propto\Sigma_{\rm gas}^{1-y}$ decreases with increasing $\Sigma_{\rm gas}$ for
$y>1$. \cite{colombo18} determined the ratio of the gas depletion time to the orbit
time for 39 local galaxies, finding a nearly constant ratio within each Hubble type
and a systematically smaller ratio for later types.  This result is consistent with
our decrease in $t_{\rm dep}$, as orbit time decreases closer to the center where
$\Sigma_{\rm mol}$ is increasing.

Finally, we calculate the star formation efficiency per free-fall time,
$\epsilon_{\rm ff}$ from
\begin{equation}
\epsilon_{\rm ff} = t_{\rm ff} {\Sigma_{\rm SFR}\over \Sigma_{\rm mol}}
= {\sqrt{3}\over{4 G}}
 {\sigma_v \Sigma_{\rm SFR}\over \Sigma_{\rm mol}^2}
\label{eqepsilon}
\end{equation}
Figure~\ref{fig:3panel} shows that $\epsilon_{\rm ff}$ is roughly constant with mean
values of 5-7\% in each of the 5 galaxies. These efficiencies are nearly an order of
magnitude larger than those determined in spiral disks \citep[e.g.,][]{utomo2018},
and similar to the efficiencies estimated for individual molecular clouds \citep[see
review in][]{krumholz18}.
The variable X(CO) from
Eq. \ref{eq:xco} would result in a trend of increasing efficiency
with increasing gas surface density.

\section{Discussion}
\label{sec:disc}

There are several models for star formation and the origin of interstellar
turbulence that make predictions about how the various quantities discussed here
should scale with each other. These models usually involve some feedback control
involving these quantities and the star formation rate, and they all reproduce the
KS relation well enough. An important question is whether they also reproduce the
other relations found here, such as the correlation of velocity
dispersion with gas surface density.

There are essentially four feedback processes that seem to be important for
interstellar equilibria: (1) the regulation of a 2-phase interstellar medium through heating by
starlight; (2) the limitation of density and collapse rate in molecular clouds
through dispersal by internal star formation; (3) the maintenance of the disk scale
height, midplane density, and average star formation rate through
turbulence driven
by star formation; and (4) the maintenance of a marginally stable
interstellar medium on large
scales through self-control of the Toomre $Q$ parameter. 
(Feedback driven by an AGN is beyond the
scope of this discussion.)
We view the first of these as a minimum constraint for star formation to occur at
all, since gaseous gravity needs cool and moderate-density clouds to bring together
in order to make the giant molecular clouds in which most stars form. A recent
observational confirmation of this type of thermal feedback is in
\cite{herrera-camus17}, and a detailed model is in \cite{hill18}.
The second and third processes control the star formation rate by different means,
the second on small scales inside individual OB associations by clearing away
the dense gas and stopping star formation locally, and the third on scales
comparable to the scale height by pumping interstellar turbulence and inflating the
disk so as to lower the average density and slow the large-scale collapse. The
fourth process controls the velocity dispersion of the gas through spiral and
large-scale Jeans instabilities, which operate faster and pump in more turbulent
energy when the dispersion is low. These Jeans instabilities may also promote giant
molecular cloud formation, giving the fourth process also a connection to star
formation.

\cite{ostriker10} proposed a model that is applicable to starburst galaxies like
those considered here, although their simulations probe somewhat lower
gas surface densities of 100-1000 $M_\odot$ pc$^{-2}$. Along with \cite{ostriker11} and \cite{shetty12}, they
suggest that the star formation rate and the phases of the interstellar medium 
are both regulated by massive young stars through momentum input via supernovae and
radiative heating, respectively. With momentum input primarily from supernovae
(which dominate stellar winds, HII regions, and radiative forcing on grains -- see
\citealt{ostriker11}), the turbulent speed equals approximately $0.4\epsilon_{\rm
ff} p^*/m^*$, where $\epsilon_{\rm ff}\sim0.005$ is the assumed efficiency of star formation
per unit free fall time at the midplane density, and $p^*/m^*\sim3000$ km s$^{-1}$
is the assumed supernova momentum input per unit stellar mass formed.  The
velocity dispersion derived in this way equals a fixed $\sim6$ km s$^{-1}$,
independent of $\Sigma_{\rm gas}$ or $\Sigma_{\rm SFR}$ \citep[see equation 22
in][]{ostriker11}. Numerical simulations in a shearing box that resolve the disk
thickness \citep{shetty12} confirm this result, showing that $\sigma$ increases from
only 4 to 5 km s$^{-1}$ as $\Sigma_{\rm gas}$ increases by a factor of $\sim10$ (see
their figure 11).  As a result of this near constancy in velocity dispersion, the
predicted disk scale height varies inversely with $\Sigma_{\rm gas}$.
Their simulations show this inverse relationship \citep[figure 13a in][]{shetty12}
over a factor of $\sim10$ in $\Sigma_{\rm gas}$, but the trend in $H$
is flattened somewhat by a
corresponding increase in the ratio of turbulent pressure to vertical momentum flux
from star formation. These predictions differ from the observations here which show
a $\sigma_v$ that increases with $\Sigma_{\rm gas}$ and imply a
constant $H$.

 The U/LIRGS in our sample have more important sources of turbulence than
supernova and stellar feedback, such as gas accretion and large-scale shocks and
tidal forces from a merger. For example, HI velocity dispersions are $\sim5\times$
higher than normal in the interacting galaxies NGC 2207/IC2163 \citep{elmegreen93},
Arp 82 \citep{kaufman97}, Arp 84 \citep{kaufman99}, and NGC 5774/5 \citep{irwin94}. A
model for these increases was in \cite{wetzstein07}. Tidal forcing of turbulence was
also proposed for the Small Magellanic Cloud by \cite{chepurnov15}.

Regarding the fourth feedback process mentioned above, observations of the
multi-fluid stability parameter $Q_{\rm 3F}$, including stars, atomic gas and
molecular gas, find that $Q_{\rm 3F}$ is about constant for all measured radii in a
large number of galaxies in the HERACLES and THINGS surveys
\citep{romeo17}. These
observations also suggest that this marginal stability is regulated mostly by the
stellar component, in which case $\sigma_v$ for the gas results from kinetic energy
input through stellar gravitational processes, such as spiral waves and spiral shock
fronts. Such a result was also demonstrated numerically in simulations by
\cite{bournaud10} and \cite{combes12} for whole galaxy disks. Those simulations
reproduced the whole-disk power spectra observed in the LMC and M33, respectively,
including the transition from a relatively flat power spectrum on large scales to a
steeper power spectrum on small scales, with the break scale equal to the disk
thickness \citep[see also][]{elmegreen01}. What is important for the present
discussion is that this double power-law power spectrum arose in simulations
both with and without star formation feedback, suggesting that even the 3D part of
the turbulence, on scales smaller than the disk thickness, can arise entirely from a
turbulent cascade from larger-scale 2D turbulence driven by disk gravity. The
primary role of supernova and other young stellar feedback in these models was to
break apart the dense clouds that form, preventing too much dense gas and too much
star formation (i.e., the second process mentioned above). Models in
\cite{hopkins11}, \cite{orr18} and others in the FIRE simulation group also stress
the importance of cloud-dispersing feedback to prevent too much star formation. This
mode of young stellar feedback, which operates on the scale of giant molecular
clouds, is distinct from that in \cite{ostriker10}, which is proposed to operate on
the scale of the disk thickness.


The cascade from large-scale 2D motions driven by disk self-gravity to small scale
3D motions, including vertical motions that affect the scale height, was illustrated
in \cite{shi14}. They show in their figure 9 how radial motions induced by disk
gravity converge on a point and mix at high pressure, diverting some of the kinetic
energy into the vertical direction.  Their figure 7 shows an increase in the
velocity dispersion with vertical position in the disk.  \cite{shi14} note that
these high latitudes are not unstable by themselves but are forced to be turbulent
by long-range gravitational forces from mass perturbations centered on the midplane.
Observations of the variation of gas velocity dispersion with height in a galaxy
might distinguish between gravity-driven turbulence and stellar-feedback driven
turbulence on the scale of the disk thickness.

\cite{bournaud09} show simulation results for self-gravitating disks that are more
directly related to the observations here. Their figure 4 plots the disk thickness
versus radius in six model galaxies where turbulent forcing is entirely by disk
gravity. The thickness is constant because the forcing by disk mass perturbations is
proportional to the inertial response by the same mass. They compare this to
external forcing by minor mergers, which produces a disk flare.  The constant
thicknesses of the starburst disks observed here could result from the same internal
gravitational forcing, with star formation feedback playing a more local role in
preventing excessively high gas densities and run-away star formation.

This discussion illustrates some of the complexities involved with feedback,
including the many types of feedback. Sometimes an observation can support two or
more physically distinct models. The KS relation is such an observation because
it contains only the projected star formation rate and gas surface
density and does not include 
additional information such as the velocity dispersion, which might
be used to distinguish among the theories. A recent observation of
star formation
rates in gas-rich starburst galaxies similar to those discussed here illustrates the ambiguity.
\citet{fisher19} derive star formation rates, surface densities, pressures, and other
quantities to test a feedback model of star formation. One result was that
$\Sigma_{\rm SFR}\propto P^{0.75}$ for pressure $P$, which is the same as the usual
KS relation, $\Sigma_{\rm SFR}\propto \Sigma_{\rm gas}^{1.5}$ if $P\propto G
\Sigma_{\rm gas}^2$ when stars and gas scale together, as they assume. Thus, we
suspect their galaxies also have a constant disk thickness.
\citep{fisher19} find 
\begin{equation}
\log(t_{\rm dep}) = -1.04 \log(\sigma) + 1.71
\label{eq:fisher}
\end{equation}
for $t_{\rm dep}$ in Gyr and velocity dispersion $\sigma_v$ in km s$^{-1}$. If we let
\begin{equation}
t_{\rm dep} = t_{\rm ff}/\epsilon_{\rm ff} = \left({{3\pi}\over{32 G \rho}}\right)^{1/2}/\epsilon_{\rm ff}
\end{equation}
and use equation (\ref{eq:fisher}) to substitute $\sigma_v$ for $t_{\rm dep}$ along
with $\epsilon_{\rm ff}=0.01$, we obtain
\begin{equation}
\sigma_v =7.7\times10^{17} \rho^{1/2}\;{\rm (cgs \;units)}.
\end{equation}
But from equation (\ref{eqH}),
\begin{equation}
H = 0.5{{\sigma_v^2}\over{\pi G \Sigma_{\rm gas}}} =
2.97\times10^{35} {{\rho}\over{\pi G \Sigma_{\rm gas}}} = 2.97\times10^{35}
\left({{\Sigma_{\rm gas}}\over{2H\pi G\Sigma_{\rm gas}}}\right)=7.08\times10^{41}/H
\end{equation}
from which we can multiply both sides by $H$, take the square root, and convert to
pc to obtain a constant (and relatively large) $H = 270$~pc. (A higher
efficiency of $\epsilon_{\rm ff}=0.05$ would result in $H = 54$~pc.)
Thus, the \cite{fisher19} result suggests a constant $H$, which is unlike
their preferred feedback model of \cite{ostriker10}.
This is not to say there is no star formation
feedback, but that $H$ may come from other types of kinetic energy input, and star
formation feedback is too small-scale to dominate the turbulence that maintains $H$
in our galaxies.

Other recent studies of the origin of interstellar turbulence have varying
conclusions.  \cite{zhou17} observed 8 local galaxies and found no correlation
between the turbulent speed and star formation rate per unit area. They also found
that the turbulent speed was higher than what was expected from star formation alone
and that additional sources are needed such as self-gravity, shear and magnetic
instabilities. \cite{johnson18} observed several hundred star-forming galaxies at
small and intermediate redshifts and found a slight increasing trend of velocity
dispersion with star formation rate density, but also suggested an important role
for gravitational instabilities at high gas fractions in driving turbulence. Still,
they could not distinguish between models where turbulence is driven by star
formation feedback from those where turbulence is driven by self-gravity.
\cite{hung19} model the turbulent speed as a function of cosmological redshift,
including accretion, star formation feedback and disk self-gravity. They find that
star formation bursts follow accretion bursts, but so does star formation feedback,
and all three sources contribute to turbulence in different degrees at different
times. \cite{yu19} also find that a combination of young stellar feedback and
gravitational instabilities are required for the observed turbulence. On smaller
scales, \cite{jin17} found in simulations that turbulence in Milky Way type
molecular clouds can be a remnant of their formation by large-scale gravitational
instabilities, without needing star formation feedback. Similarly, \cite{vazquez19}
suggest that molecular cloud motions are gravitational in origin, although more like
collapse than turbulence.

\section{Conclusions} \label{sec:conc}

A new analysis of ALMA  data for 5 luminous and ultra-luminous infrared galaxies indicates that
the KS relation for $\Sigma_{\rm gas}>10^3\;M_\odot$~pc$^{-2}$ has a slope of
$1.74^{+0.09}_{\rm -0.07}$, slightly steeper than the slope of the KS relation for
total gas in main galaxy disks.
Combining the molecular surface densities and velocity dispersions, we determine
empirical gas scale heights of 150-190~pc, with little systematic variation over 1.5
orders of magnitude in $\Sigma_{\rm gas}$. This nearly constant scale height implies
that the average midplane density varies almost linearly with the gas surface
density, and thus the gas dynamical rate varies approximately with the square root
of this surface density, giving the observed super-linear slope in the KS relation.

Star formation appears to be initiated by gravitational condensations in the average
interstellar medium, with a rate given by the free-fall time obtained
from the square root of average density and the
average density given by the ratio of the surface density to the disk thickness.
Turbulent speeds that determine this thickness may also come from gravitational
energy, pumped in by spiral arms and the associated shocks and by kpc-scale Jeans
instabilities. Star formation feedback then plays the essential role of halting the
collapse at a high density so as to prevent all the gas from turning into stars in a
free-fall time. In this interpretation, star formation feedback helps regulate the efficiency per
unit free fall time by partitioning the gas into a wide range of densities and
dispersing the densest regions before they collapse completely.

Averaged over sufficiently large scales, the star formation rate is well
approximated by the three-dimensional dynamical law, $\epsilon_{\rm ff}\rho_{\rm
mid}/t_{\rm ff}$, for free fall time $t_{\rm ff}$ proportional to $(G\rho_{\rm
mid})^{-0.5}$ and an approximately constant efficiency per unit free fall time,
$\epsilon_{\rm ff}$. Because the disk thickness and $\epsilon_{\rm ff}$ are much
more constant than either $\Sigma_{\rm gas}$ or $\rho_{\rm mid}$, integration over
the thickness of the disk in starburst galaxies preserves the mathematical form of
this law with the substitution of surface density for space density. This model is
roughly similar in spiral galaxies, although it may be more complicated in the
gas-dominated regions if the scale height varies significantly
\citep[e.g.,][]{barnes12,elmegreen15a,elmegreen15,bacchini2019}.

The reasons for the nearly constant values of $\epsilon_{\rm ff}$ and scale height
$H$ in these starbursts are not evident from the present observations although
several possibilities were suggested in Section \ref{sec:disc}. Both are larger than
in disk galaxies with lower star formation rates: $\epsilon_{\rm ff}$ by a factor of
$\sim 5-10$ and $H$ by a factor of $\sim 2$. However, these higher values barely
affect the overall KS relation because the gas surface density and star formation
rate per unit area vary by many orders of magnitude, 
and the higher $\epsilon_{\rm ff}$ and $H$ found
here partially cancel each other in the three-dimensional dynamical law of star
formation.

\acknowledgments

Acknowledgements: We are grateful to Dr. J.M.D. Kruijssen and the organizers of the
conference ``The Multi-Scale Physics of Star Formation and Feedback during Galaxy
Formation'' in Heidelberg, June 2018, for providing a collaborative atmosphere where
two of the authors, B.G.E. and C.D.W., began discussing this project. We also
acknowledge useful comments by the referee. C.D.W. acknowledges financial support
from the Canada Council for the Arts through a Killam Research Fellowship. The
research of C.D.W. is supported by grants from the Natural Sciences and Engineering
Research Council of Canada and the Canada Research Chairs program. AB wishes to
acknowledge partial support from an Ontario Trillium Scholarship (OTS). This paper
makes use of the following ALMA data: ADS/JAO.ALMA\#2011.0.00525.S,
ADS/JAO.ALMA\#2012.1.00165.S, ADS/JAO.ALMA\#2013.1.00218.S,
ADS/JAO.ALMA\#2013.1.00379.S, ADS/JAO.ALMA\#2015.1.00167.S, \hfill\break
ADS/JAO.ALMA\#2015.1.00287.S, ADS/JAO.ALMA\#2015.1.00993.S,
ADS/JAO.ALMA\#2016.1.00140.S.
 ALMA is a partnership of ESO
(representing its member states), NSF (USA) and NINS (Japan), together
with NRC (Canada), MOST and ASIAA (Taiwan), and KASI (Republic of
Korea), in cooperation with the Republic of Chile. The Joint ALMA
Observatory is operated by ESO, AUI/NRAO and NAOJ. The National Radio
Astronomy Observatory is a facility of the National Science Foundation
operated under cooperative agreement by Associated Universities, Inc.
This research has made use of the NASA/IPAC Extragalactic Database
(NED), which is operated by the Jet Propulsion Laboratory, California
Institute of Technology, under contract with the National Aeronautics
and Space Administration.

\vspace{5mm}
\facilities{ALMA}


\software{astropy \citep{2013A&A...558A..33A},
          CASA \citep{mcmullin2007}
          }

\appendix

\section{Binned Data Table}

We provide the binned data used in this paper as a machine-readable
table. In this table, each row reports our measurements for one large
pixel in one galaxy. The contents of the rows are as follows:

\begin{enumerate}

\item The name of the galaxy tagged with a pixel identifier, and the
  central coordinates of the pixel in decimal degrees;

\item The observed CO J=1-0 
  CO velocity dispersion with its corresponding rms uncertainty;

\item The logarithmic value and logarithmic uncertainty for: 
the
  molecular gas surface density, $\Sigma_{\rm mol}$; the star
  formation rate surface density, $\Sigma_{\rm SFR}$; the
  depletion time, $t_{\rm dep}$; the
  free-fall time, $t_{\rm ff}$; the efficiency per free-fall time,
  $\epsilon_{\rm ff}$; and the molecular gas scale height, $H$.

\end{enumerate}

\end{document}